\documentclass[preprint]{aastex}
\usepackage{graphicx}
\slugcomment{To appear in IAU Symposium No.~273}

\begin{document}
%\begin{document}
\title{Long-term Evolution of Sunspot Magnetic Fields}
\author{Matthew J. Penn \and William Livingston}
\affil{National Solar Observatory\altaffilmark{1}, 950 N Cherry Av, Tucson AZ 85718}
\email {mpenn@nso.edu}

\altaffiltext{1}{NSO is operated by AURA, Inc., under contract to the
National Science Foundation.}

%\maketitle

\begin{abstract}
Independent of the normal solar cycle,
a decrease in the sunspot magnetic field strength has been observed
using the Zeeman-split 1564.8nm Fe I spectral line
at the NSO Kitt Peak McMath-Pierce telescope.
Corresponding changes in sunspot brightness and the strength of
molecular absorption lines were also seen.
This trend was seen to continue in observations of the first sunspots
of the new solar Cycle 24,
and extrapolating a linear fit to this trend would lead to only half the number
of spots in Cycle 24 compared to Cycle 23,
and imply virtually no sunspots in Cycle 25.

We examined synoptic observations from the NSO Kitt Peak Vacuum Telescope
and initially (with 4000 spots) found a change in sunspot brightness
which roughly agreed with the infrared observations.
A more detailed examination (with 13,000 spots)
of both spot brightness and line-of-sight magnetic flux
reveals that the relationship of the sunspot magnetic fields
with spot brightness and size remain constant during the solar cycle.
There are only small temporal variations in the
spot brightness, size, and line-of-sight flux seen in this larger sample.
Because of the apparent disagreement between the two data sets,
we discuss how the infrared spectral line provides a uniquely direct
measurement of the magnetic fields in sunspots.
%\keywords{}
%% add here a maximum of 10 keywords, to be taken form the file <Keywords.txt>
\end{abstract}

%\firstsection % if your document starts with a section,
%              % remove some space above using this command.
\section{Introduction}

Observations of the magnetic fields in sunspot umbrae have been carried out by
Livingston at the National Solar Observatory's McMath-Pierce solar telescope atop
Kitt Peak.
These observations are made with a single-element detector and measure the
intensity spectra of the 1564.8nm Fe I g=3 spectral line and
nearby atomic and molecular absorption lines.
While these observations began in the 1990's, the focus then was only on the larger sunspots
visible on the solar disk.
During the last 10 years these observations have become more synoptic in that all
sunspots visible on the solar disk are observed in this way,
from solar pores to the 
largest umbrae.
(In the following text we use the term "spots"
to refer to both sunspots with penumbrae and pores without penumbrae.)
After fitting several spectral lines in the data, Livingston has compiled a table
of the magnetic field strength at the darkest spot location, the continuum
brightness at that location (normalized to nearby quiet Sun brightness), and the 
line depth of several OH molecular lines in the spectral field-of-view.
It is important here to note
(1) that no polarimetry is done, only intensity spectra are used,
(2) that the 1564.8nm Fe I line is completely split
(i.e. the Zeeman sigma components are shifted in wavelength more than their line widths)
for the 1500 Gauss and larger magnetic fields seen in the spots,
and
(3) the splitting of the sigma components in the intensity spectrum measures the
true magnetic field strength, not a vector component of the magnetic field.

We reported in
%\cite[Penn \& Livingston (2006)]{PennLiv2006}
\citet{PennLiv2006}
that a time series of this magnetic field data showed 
a decrease in the umbral magnetic field strength which was independent of the normal
sunspot cycle.
Also, the measurements revealed a threshold magnetic field strength of about 1500 Gauss,
below which no dark pores formed.
A linear extrapolation of the magnetic field trend suggested that the mean field
strength would reach this threshold 1500 Gauss value in the year 2017.
Furthermore, analysis of the umbral continuum brightness showed another linear trend,
and extrapolation showed the umbral brightness would be equal to the quiet Sun 
brightness at about the same year.
Finally, the molecular line depths showed a decreasing strength with time, and again
the trend suggested that molecular absorption lines would disappear from the average
sunspot umbra near 2017.

Recent observations spanning from the solar interior to the solar corona
clearly show that solar Cycle 24 has started.
Below the solar surface,
helioseismic observations of the torsional oscillations have shown that the 
subsurface flow maxima migrated to latitudes of +/-23 degrees in February of 2009
coinciding 
with the flow latitude at the onset of the magnetic activity for solar Cycle 23
%(\cite[Howe et al. 2009]{Howeetal2009}).
\citep{Howeetal2009}.
At the solar surface,
the sunspot number is rising
(http://sidc.oma.be/sunspot-data/).
The magnetic polarity of solar magnetic active regions has switched since Cycle 23,
and the hemispheres now show new cycle magnetic flux consistent with Hale's polarity law
\newline(http://www.nso.edu/press/cycle24.html).
In the solar chromosphere
the spectral Ca K index has shown an increase
(ftp://ftp.nso.edu/idl/cak\_plot.gif).
And in the solar corona,
the radio emission from the Sun at 10.7~cm wavelength has begun to increase
(http://www.spaceweather.gc.ca/sx-6-eng.php),
and the UV emission from the Sun has started to rise
\newline(http://lasp.colorado.edu/lisird/sorce/sorce\_ssi/ts.html).
And finally of note, the Solar Cycle 24 Prediction Panel
from the Space Weather Prediction Center
has recognized that the minimum after solar Cycle 23 was reached in December 2008
(http://www.swpc.noaa.gov/SolarCycle/).

If Cycle 24 has started, we are in the rise phase of the cycle;
but where exactly in the cycle are we located?
The helioseismic observations can tell us based on the latitude of
the torsional oscillation bands.
This gives us a phase indicator which is independent of 
the cycle duration or the amplitude of the activity peak for the cycle.
We can extrapolate the latitudinal drift of the torsional bands 
%(\cite[Howe et al. 2009]{Howeetal2009})
\citep{Howeetal2009}
and then compare the current position with the position in Cycle 23.
This calculation tells us that June 2010
in Cycle 24 corresponds with February 1998
in Cycle 23.
It is instructive to examine the monthly sunspot numbers for those two
months; for February 1998 that value was 40,
and for June 2010 that value was 13
(http://sidc.oma.be/sunspot-data/).
Including the 5 months preceding these times, we find that for a 6 month
period Cycle 24 has shown only 0.37 times the number of spots seen in Cycle 23.
By correcting for the phase of the solar cycles,
we are now seeing far fewer sunspots than we saw in the preceding cycle;
solar Cycle 24 is producing an anomalously low number of dark spots and pores.

\section{Recent Observations}

\begin{figure}[htb]
% \vspace*{-2.0 cm}
\begin{center}
 \includegraphics[width=5.0in]{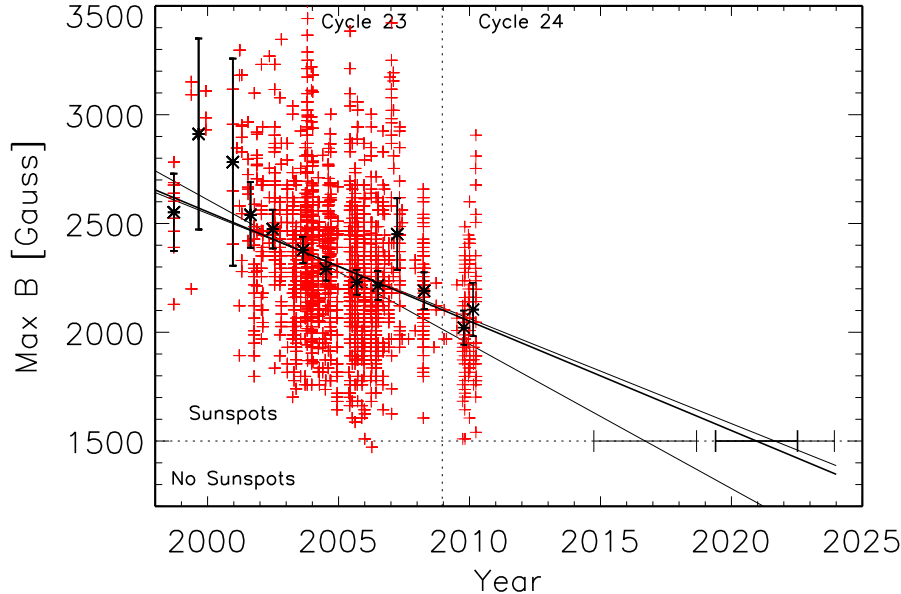} 
% \vspace*{-1.0 cm}
 \caption{Measurements of the total magnetic field strength at the darkest location in 
  umbrae and pores as a function of time.
  The crosses show the individual measurements,
  the asterisks show annual bins.
  Three linear fits are shown: the bottom fit line fits data from 1998-2006 as 
  done in our 2006 paper.
  The top line fits all the data from Cycle 23, and the middle
  line fits all of the data.}
   \label{fig1}
\end{center}
\end{figure}

Figure\,\ref{fig1}
shows the observations of sunspot and pore magnetic fields from Livingston's data set.
The total magnetic field strength at the darkest location
in the umbra or pore is plotted against the date of the measurement.
The raw measurements are shown as crosses.
There is a large distribution of magnetic field strengths in spots visible 
on the solar photosphere, and there seems to be a lower threshold for the
formation of dark spots, either pores or umbrae.
No measurements show that the total magnetic field strength
is less than about 1500 Gauss in a dark spot,
and presumably magnetic regions with maximal field strengths less than this 
value do not undergo convective collapse.
In Figure~1 the annual bins of the measurements are shown as asterisks,
and the standard deviation of the mean is shown as a vertical error bar on the asterisks.

Various linear fits are also shown in the Figure.
The line to the left shows a linear fit from the work done by 
%\cite[Penn \& Livingston (2006)]{PennLiv2006};
\citet{PennLiv2006};
the extrapolated line shows an intercept with the 1500 Gauss value in
2017, and error bars of the computed intercept are also shown.
The right-most line fits all of the data from Livingston's Cycle 23 observations, 
and the slight uptick in the magnetic field measurements from 2007 and 2008
move the 1500 Gauss intercept time out to 2022.
The central line fits all of the data, including measurements from Cycle 24,
and the intercept date now appears to be 2021, but it is within the
error-bars from the fit to the Cycle 23 sunspot data.
The linear fit to all of the data show a decrease of about 50 Gauss per year
in the magnetic field strength at the darkest location in spots.

It is important to note that both sunspots and pores are included in this plot.
Pores, lacking penumbra, often have magnetic fields less than 2000 Gauss,
but always have magnetic fields stronger than 1500 Gauss.
Secondly, the intercept of the mean magnetic field strength
with this 1500 Gauss threshold
does not imply 
that all sunspots will disappear by the year 2021; 
rather it implies that half of the sunspots which would normally appear 
on the surface of the Sun would be visible.
Finally, the plot doesn't address the other magnetic fields on the Sun where
field strengths are lower than 1500 Gauss;
the temporal behavior of solar active network or quiet Sun magnetic fields
may be different from the behavior shown by sunspots.

\section{Searching for support in other Data}

The changes observed in sunspot brightness prompted an analysis of
sunspot umbrae as observed in the synoptic data set from the
National Solar Observatory Kitt Peak Vacuum Telescope (KPVT).
In the first analysis of that data set, 
%\cite[Penn \& MacDonald (2007)]{PennMac2007}
\citet{PennMac2007}
selected isolated spots by hand.
About 4000 sunspots and pores were examined, and a cyclic behavior was
seen in the minimum brightness found in these dark spots in phase with the 
sunspot cycle; darker sunspots were more common during solar maximum,
and brighter sunspots were more common during solar minimum.
Strangely, no significant temporal change in the radius of the sunspot 
umbrae was seen.
Using well-known wavelength scaling coefficients from Maltby
%(\cite[Maltby et al. (1986)]{Maltbyetal1986})
\citep{Maltbyetal1986}
the KPVT data showed a good correspondence with both a study 
using MDI data
%(\cite[Norton \& Gilman (2004)]{NortonGilman2004})
\citep{NortonGilman2004}
and with the observations of sunspot intensities in the infrared
by Livingston.
At the time, the uptick in the magnetic field strengths seen by
Livingston in 2007 and 2008 suggested that perhaps there was a solar-cycle
dependence.

A more detailed analysis of the KPVT data set was performed by Tom Schad
%(\cite[Schad \& Penn (2010)]{SchadPenn2010})
\citep{SchadPenn2010}
which included an automated sunspot selection procedure,
resulting in the identification of over 13,000 dark spots,
and an analysis of the brightness as well as the line-of-sight magnetogram data.
This work showed that there were only small temporal changes in the 
spot intensities and magnetic field strengths.
It also showed that two empirical sunspot relationships,
the first between sunspot magnetic field strength and brightness,
and the second between magnetic field strength and spot radius,
both remained unchanged during the solar cycle.
Both relationships did contain some scatter, but it was found that
the temporal changes in the spot radius were consistent with the changes 
in magnetic fields and brightness.
Finally, current work with the KPVT data set suggests that the twist of the 
sunspot magnetic fields does not vary significantly during the solar cycle.
The horizontal pressure balance that spots achieve with the surrounding
quiet Sun behaves the same way at all phases of the solar cycle.

Work from other authors have addressed some of these issues as well.
Observations of the brightness of sunspots as measured with MDI showed
no changes from 1998-2004
%(\cite[Mathew et al. (2007)]{Mathewetal2007})
\citep{Mathewetal2007}
which is consistent with the observed KPVT data during this time interval.
Measurements of the brightness of sunspot umbrae from
the California State University Northridge
San Fernando Observatory showed no changes during the interval from
1997-2004
%(\cite[Wesolowski, Walton \& Chapman (2008)]{Wesolowskietal2008})
\citep{Wesolowskietal2008}
although the brightness vs radius relationship from that data seems anomalous
%(\cite[Schad \& Penn (2010)]{SchadPenn2010}).
\citep{SchadPenn2010}.
And most recently in these proceedings, measurements of the magnetic fields from
sunspot umbrae near the center of the solar disk using MDI magnetograms
%\cite[Watson \& Fletcher (2010)]{WatsonFletcher2010}
\citep{WatsonFletcher2010}
show a smaller decrease in the magnetic field strength,
but that result is not significant compared to the standard errors of their fit.

Measuring the true magnetic field strength in the darkest sunspot
or pore regions is known to be a difficult task since
the brightness levels are low and the line depths are small
%(\cite[Liu, Norton \& Scherrer (2007)]{Liuetal2007}).
\citep{Liuetal2007}.
Using simultaneous measurements of a large sunspot from Hinode and MDI, 
%\cite[Moon et al. (2007)]{Moonetal2010}
\citet{Moonetal2010}
show that the MDI observations can underestimate the magnetic field strength
by a factor of two.
Imaging magnetographs have distinct advantages in terms of cadence of observations
and the spatial integrity of the images,
but spectrograph-based instruments which capture full line profiles in dark spots
do have advantages in terms of accuracy.

It is also important to realize that the data obtained by Livingston using the
infrared line at 1564.8nm with a Land\'{e} g-factor of 3.0 are measuring 
magnetic fields that are completely resolved.
Using a conservative estimate (i.e. a large Doppler and instrumental line width)
the infrared spectral line can resolve fields with strengths
greater than 750 Gauss.
We can scale this value by the factor of g times lambda for many of the
instruments used to study sunspot magnetic fields.
The KPVT magnetic field resolution would be 2400 Gauss,
and for MDI the resolved field strength is about 3600 Gauss.
For the HMI and SOLIS spectrograph-based instruments, the magnetic field
strength must be above 2200 Gauss to be fully resolved.
If we examine the measurements in Figure~1 which have magnetic fields only
above 2200 Gauss, the temporal trend is not apparent.
Certainly magnetic fields can be determined for spots with fields below this
magnetic resolution value, but there are assumptions, corrections and (in some cases) models
which are used in that determination, and perhaps the scatter inherent in that
process is enough to swamp the underlying temporal variation which is so
apparent in the more direct infrared measurements.

The lack of significant brightness or radius variation of sunspots as seen 
with other instruments is more difficult to explain.
While the infrared measurements suffer from less instrumental scattered light,
and perhaps better ground-based seeing than ground-based visible observations,
these advantages do not seem large enough to explain the lack of variation
seen with other telescopes; it remains a mystery.

\section{Implications and Critical Observations for the Future}

\begin{figure}[tb]
% \vspace*{-2.0 cm}
\begin{center}
 \includegraphics[width=3.4in]{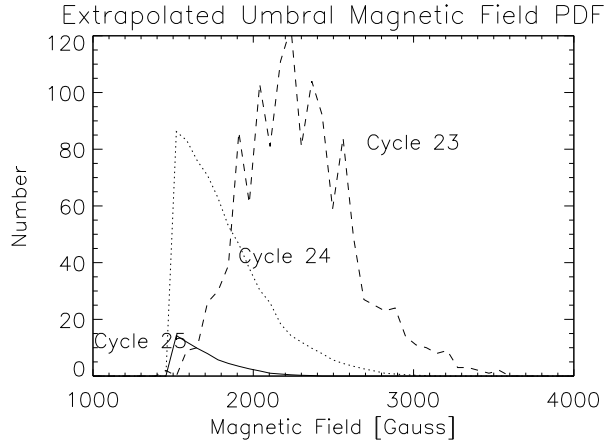} 
% \vspace*{-1.0 cm}
 \caption{The magnetic probability distribution function (PDF) is show for the IR
measurements of sunspots during Cycle 23.
With the assumption discussed in the text, we can produce PDFs for Cycles 24 and 25.
A simple scaling using the total number of spots
suggests Cycle 24 will peak with a SSN of 66, 
and Cycle 25 will peak with a SSN of 7.}
   \label{fig2}
\end{center}
\end{figure}

As suggested by Figure~1 
a detailed analysis shows that the sunspots measured during 
the rise phase of Cycle 24 have the same shape in the distribution of magnetic field strengths
as the spots seen during the 
decay phase of Cycle 23, but that the mean value of the distribution is reduced.
This is a conservative conclusion from Livingston's observations.
If we make three assumptions however, we see that there are more dramatic implications
of these infrared observations.
First we assume that the distribution of magnetic fields observed by Livingston 
from 1998-2008 is a good proxy for the true probability distribution function (PDF) for
sunspot umbral magnetic fields for Cycle 23.
Secondly, we assume that the magnetic threshold of 1500 Gauss represents a real physical limit
for the formation of a dark spot (either a pore or a sunspot) on the solar photosphere.
And finally, we assume that the mean of the magnetic field PDF continues to decrease
linearly with time.

Figure~2 shows the computed magnetic PDF for the sunspots in cycles 24 and 25,
using a linear decrease of the magnetic field of 65 Gauss per year and a duration
of 11 years for each cycle.
This is meant to represent an upper limit, and the magnetic change corresponds to the most steeply 
sloped line in Figure~1.
We can see that the PDFs for Cycle 24 and Cycle 25 vary dramatically from that
observed in Cycle 23.
If we assume that the appearance time of sunspots during each cycle is similar, we
can use the total number of spots in each cycle to compute the maximum activity
level of that cycle, using the fact that Cycle 23 showed
a peak smoothed sunspot number (SSN) of 130.
The linear decrease of 65 Gauss per year predicts that Cycle 24 will peak with a 
smoothed SSN of 66, and Cycle 25  will peak with a smoothed SSN of 7.
Using a value of 50 Gauss per year suggests
a smoothed SSN of 87 for Cycle 24 and 20 for Cycle 25.

It is important to note that it is always risky to extrapolate linear trends;
but the importance of the implications from making such an assumption justify its mention.
Also of note is that while these PDFs are drawn from Livingston's observations,
they are at best proxies for the true sunspot magnetic PDFs.
While a sunspot with 
a magnetic field strength of 4200 Gauss was observed in Cycle 23 (NOAA 10930, 
%\cite[Moon et al. (2007)]{Moonetal2010}),
\citet{Moonetal2010}),
it was not
observed by Livingston and does not appear in this analysis.
Thus the sunspot which appeared recently in Cycle 24 (NOAA 11092, August 2010)
with a magnetic field strength of 3350 Gauss does not invalidate these assumptions.
Certainly if a large number of sunspots with magnetic field
strengths greater than 3000 Gauss do appear,
then the extrapolated PDF will be shown to be erroneous.
We will see in the coming months and years.

Umbral magnetic field measurements at 1564.8nm have been shown to reveal differences
between the decay phase of Cycle 23 and the rise phase of Cycle 24,
and they imply that the next two sunspot cycles might be very different from the last one.
Observations with visible light magnetographs do not show significant
support for these claims.
Thus we feel it is essential to make synoptic observations using this very favorable
infrared line to determine if these trends continue.
It is essential to save the spectra and calibrations, and
it would be very useful to make synoptic measurements of sunspots using temperature
sensitive molecular lines such as the lines of OH near 1564.8nm.

%\begin{discussion}
%\end{discussion}


\begin{thebibliography}{}

\bibitem[Howe et al.(2009)]{Howeetal2009}
{Howe, R., Christensen-Dalsgaard, J., Hill, F., Komm, R., Schou, J. \& Thompson, M. J.}
\textit{ApJ} (Letters) 701, L87

\bibitem[Liu, Norton \& Scherrer(2007)]{Liuetal2007}
{Liu, Y., Norton, A.A. \& Scherrer, P.H.} 2007, 
\textit{Solar Physics} 241, 185 

\bibitem[Maltby et al.(1986)]{Maltbyetal1986}
{Maltby, P., Avrett, E. H., Carlsson, M., Kjeldseth-Moe, O., Kurucz, R. L., \& Loeser, R.} 1986,
\textit{ApJ} 306, 284

\bibitem[Mathew et al.(2007)]{Mathewetal2007}
{Mathew, S.K., Martinez Pillet, V., Solanki, S.K. \& Krivova, N.A.} 2007, 
\textit{Astron. Astrophys.} 465, 291 

\bibitem[Moon et al.(2007)]{Moonetal2010}
{Moon, Y.-J., Kim, Y.-H., Park, Y.-D., Ichimoto, K., Sakurai, T.,
Chae, J., Cho, K.S., Bong, S., Suematsu, Y., Tsuneta, S., Katsukawa, Y.,
Shimojo, M., Shimizu, T., Shine, R.A., Tarbell, T.D., Title, A.M., Lites, B.,
Kubo, M., Nagata, S., \& Yokoyama, T.} 2007,
\textit{PASJ} 59, 625.

\bibitem[Norton \& Gilman(2004)]{NortonGilman2004}
{Norton, A.A. \& Gilman, P.A.} 2004,
\textit{ApJ} 603, 348.

\bibitem[Penn \& Livingston(2006)]{PennLiv2006}
{Penn, M.J. \& Livingston, W.} 2006, 
\textit{ApJ} (Letters), 649, L45

\bibitem[Penn \& MacDonald(2007)]{PennMac2007}
{Penn, M.J. \& MacDonald, R.K.D.} 2007, 
\textit{ApJ} (Letters), 662, L123

\bibitem[Schad \& Penn(2010)]{SchadPenn2010}
{Schad, T.A. \& Penn, M.J.} 2010, 
\textit{Solar Physics} 262, 19 

\bibitem[Watson \& Fletcher(2010)]{WatsonFletcher2010}
{Watson, F. \& Fletcher, L.} 2010,
\textit{IAU Symposium 273, these proceedings}

\bibitem[Wesolowski, Walton \& Chapman(2008)]{Wesolowskietal2008}
{Wesolowski, M.J., Walton, S.R., Chapman, G.A.} 2008,
\textit{Solar Physics} 248, 141.

\end{thebibliography}
\end{document}